\documentclass[reprint,superscriptaddress,amsmath,amssymb,aps,prx]{revtex4-1}

\usepackage{graphicx}
\usepackage{natbib}
\usepackage[colorlinks=true,citecolor=blue]{hyperref}
\usepackage{cleveref}
\usepackage[caption=false,labelformat=empty,captionskip=0pt]{subfig}

\newcommand{\spR}{{\eta_{\omega,\Omega}}}
\newcommand{\ncr}{{n_\text{cr}}}
\newcommand{\micron}{{\mu\mathrm{m}}}
\newcommand{\Wcm}{\mathrm{Wcm}}

\newcommand{\rmd}{\mathrm{d}}
\newcommand{\fs}{\mathrm{fs}}
\newcommand{\eV}{\mathrm{eV}}
\newcommand{\aeff}{a_\text{eff}}
\newcommand{\xce}{\eta_\text{XUV}}
\newcommand{\nm}{\text{nm}}
\newcommand{\Seff}{S_\text{eff}}

\begin{document}

\title{Relativistically intense XUV radiation from laser-illuminated near-critical plasmas}

\author{T. G. Blackburn}
\email{tom.blackburn@chalmers.se}
\affiliation{Department of Physics, Chalmers University of Technology, SE-41296 Gothenburg, Sweden}
\author{A. A. Gonoskov}
\affiliation{Department of Physics, Chalmers University of Technology, SE-41296 Gothenburg, Sweden}
\affiliation{Institute of Applied Physics, Russian Academy of Sciences, Nizhny Novgorod 603950, Russia}
\affiliation{Lobachevsky State University of Nizhni Novgorod, Nizhny Novgorod 603950, Russia}
\author{M. Marklund}
\affiliation{Department of Physics, Chalmers University of Technology, SE-41296 Gothenburg, Sweden}

\date{\today}

\begin{abstract}
Pulses of extreme ultraviolet (XUV) light, with wavelengths between 10 and 100$\,$nm, can be used to image and excite ultra-fast phenomena such as the motion of atomic electrons.
Here we show that the illumination of plasma with near-critical electron density may be used as a source of relativistically intense XUV radiation, providing the means for novel XUV-pump--XUV-probe experiments in the non-linear regime.
We describe how the optimal regime may be reached by tailoring the laser-target interaction  parameters and by the presence of preplasma.
Our results indicate that currently available laser facilities are capable of producing XUV pulses with duration $\sim 10\,\fs$, brilliance in excess of $10^{23}$~photons/s/mm$^2$/mrad$^2$ (0.1\% bandwidth) and intensity $I\lambda^2 \gtrsim 10^{19}\,\Wcm^{-2}\micron^2$.
\end{abstract}


\maketitle

\section{Introduction}

There has been extensive demonstration of radiation sources in the
extreme ultraviolet (XUV) to X-ray frequency range employing laser
illumination of solid~\cite{RousseKAlpha,Dromey,Quere,Dromey2}
and gaseous targets~\cite{RousseGas,Kneip,Corde}.
High-frequency radiation naturally arises in these interactions due to
the non-linear motion of electrons oscillating in the strong
electromagnetic fields of an intense laser pulse.
The generation efficiency and spectral properties of XUV sources
are of paramount importance for applications in
diagnostic imaging~\cite{Dobosz,Brenner}, the
creation and study of warm dense matter~\cite{Nagler,Galtier}
and for probing phenomena at the attosecond~\cite{Krausz}
and femtosecond scales~\cite{Tzallas}.

Here we show that the XUV pulses emitted by laser-illuminated near-critical
plasmas can be relativistically intense, i.e. they are strong enough to
accelerate electrons from rest to relativistic velocity in a single cycle.
The physical origin of this emission is the near-micron scale oscillation of the
plasma surface, which leads to the storage and re-emission of a large
fraction of the laser energy once per optical cycle.
This ultimately compresses the re-emission into an attosecond burst that has larger
electric-field amplitude than the incident light~\cite{Brugge,Gonoskov}.
The generation process is optimal when plasma with electron density
between 1 and 4 times the relativistic critical density is irradiated
at $60^\circ$ to the target normal. Under these conditions, 5\% of the
laser energy is converted to harmonics in the XUV frequency range and
the re-emitted pulse has peak intensity four times larger than the
incident pulse. This mechanism is robust against the presence of preplasma
and can, in fact, be enabled by it.

Laser-irradiated solid-density plasmas have attracted attention as
sources of XUV light as, unlike gas-based sources, there is no upper limit
on the input intensity~\cite{TeubnerReview}.
For intensities $\sim10^{16}\,\Wcm^{-2}$, below the relativistic threshold,
coherent wake emission leads to harmonic generation in the
ultraviolet~\cite{Quere}; at higher intensities, experiments have demonstrated
conversion efficiencies of 0.01 ($10^{-5}$) to X-rays with energy greater
than 20~eV (1~keV)~\cite{DromeyPRL}, or $10^{-4}$ in the tens of eV~\cite{Rodel}
at $\sim10^{19}\,\Wcm^{-2}$.
The source of these high harmonics is the collective motion of electrons at the
illuminated plasma surface, rather than atomic ionization and recombination.

Harmonic generation is commonly described by the \emph{relativistic
oscillating mirror} (ROM) model, in which the incident light is
reflected at every instant of time within the wave cycle from a certain
oscillating point~\cite{Bulanov,Lichters}. The Doppler shift when this
point moves back towards the observer at relativistic velocity leads to the
increase in frequency of the reflected light. Models based on this
assumption have given insight into polarization selection
rules~\cite{Lichters,VonDerLinde}, the angular dependence of generation
efficiency~\cite{Debayle}, and the power-law form of the intensity
spectrum~\cite{Baeva,Pirozhkov,Debayle2,Boyd,Boyd2}.

The descriptive power of the ROM model as formulated in \cite{Baeva} arises from the assumption of
Leontovich boundary conditions at the oscillating surface, which imply
instantaneous equality between incoming and outgoing energy flux and
therefore bound the electric-field amplitude of the reflected light to that of the
incident light. For highly-overdense plasmas with
steep density profile this assumption and its associated models work well.

However, with either increase of incident electric-field amplitude or decrease of plasma
density, the assumption of instantaneous reflection begins to break down.
Instead energy is first accumulated in a transient charge separation
field when the laser radiation pressure displaces electrons from the
plasma-vacuum boundary. Due to the relativistic nature of the motion,
the displaced electrons are compressed into a thin layer, which re-emits
the stored energy when it propagates back towards the boundary.
The parameter regime where this occurs provides an opportunity not only
to compress the pulse in time but also to increase the re-emitted intensity
above that of the incident wave.

The \emph{relativistic electron spring} (RES) model~\cite{Gonoskov,Gonoskov2}
describes these interactions by assuming that emission from
the separated ions and electrons compensates the incident radiation within
the plasma bulk, while allowing energy accumulation in the separation region.
This assumption is weaker than that of instantaneous reflection, allowing
for the description of a broader set of dynamics. The key result is that the
generated radiation may have significantly higher electric-field amplitude than the
incident light.
In the ROM regime by contrast, the electric-field amplitude is
strictly constrained to that of the incident light.

The RES model describes the plasma microdynamics in terms of the instantaneous
displacement and velocity of a thin electron layer, the current of which
generates coherent synchrotron emission (CSE). This
has been studied in the context of creating reflected and transmitted
attosecond pulses~\cite{DromeyCSE,Ma,Cousens,Zhang2017} and in thin-foil
interactions~\cite{Mikhailova,Bulanov2}. In the RES regime the layer
forms automatically, leading to the emission of attosecond~\cite{Bashinov}
bursts with higher electric-field amplitude of the incident light~\cite{Brugge,Gonoskov,Fuchs},
controllable ellipticity~\cite{Blanco}, as well as bright incoherent beams~\cite{Serebryakov}.
We will
show in this work that the RES equations can be used to model the plasma dynamics
and emission properties in the case that the plasma-vacuum interface
is not perfectly sharp.

There has been considerable work devoted to optimizing XUV
generation~\cite{Sansone}, including PIC simulation of few-cycle
lasers~\cite{Tsakiris,Cousens}, development of analytics~\cite{Gonoskov,Debayle,Gonoskov2}
and experiment. It has been shown, for example, that the denting of
the plasma surface by the laser light pressure, which would otherwise
increase the divergence of the high harmonics, may be compensated by
tailoring the input laser pulse~\cite{Vincenti}. Simulations have
also demonstrated the advantages of multiple reflection geometries~\cite{Edwards2}
and specially-designed waveforms for the incident light~\cite{Edwards}.
	
We use three dimensionless parameters to characterize the laser-plasma
interaction: $a_0 = e E/(m c \omega)$, the normalized amplitude of
an electromagnetic wave with electric field strength $E$ and angular
frequency $\omega$; $\theta$, the angle between the laser wavevector
and the target normal; and $S$, the ratio of the electron number
density $n_e$ to that of the relativistic critical density
$a_0 \ncr$~\cite{GordienkoS}.
Here $\ncr = \epsilon_0 m \omega^2/e^2$ is the non-relativistic
critical density, and $e$, $m$ are the
electron charge and mass, $c$ the speed of light, with $\epsilon_0$
the vacuum permittivity.
The laser peak intensity $I_0 = \tfrac{1}{2} a_0^2 m c^3 \ncr$,
or $I_0 [10^{18}\,\Wcm^{-2}] = 1.37 a_0^2 / (\lambda [\micron])^2$
for wavelength $\lambda$.
A convenient approximation for $S$ accurate within 5\% is
	\begin{equation}
	S =	\frac{n_e}{a_0 \ncr} \simeq \frac{n_{23} \lambda_\micron}{\sqrt{I_{22}}}
	\label{eq:S}
	\end{equation}
where $n_{23}$ is the electron number density in units of $10^{23}~\text{cm}^{-3}$,
$I_{22}$ is the laser intensity in units of $10^{22}~\Wcm^{-2}$ and
$\lambda_\micron$ its wavelength in microns.

Here $S$ is defined in terms of the bulk electron density of the plasma.
However, all high-intensity laser systems exhibit finite contrast, leading to
heating and expansion of the target by prepulse and pedestal light
before the arrival of the main pulse. In this work we consider such
effects by including a density ramp of scale-length $L$ in front of
the plasma bulk. We show that an effective $S$ parameter may be usefully
defined in this scenario in \cref{eq:EffectiveS}.

	\begin{figure}
	\centering
	\includegraphics[width=0.8\linewidth]{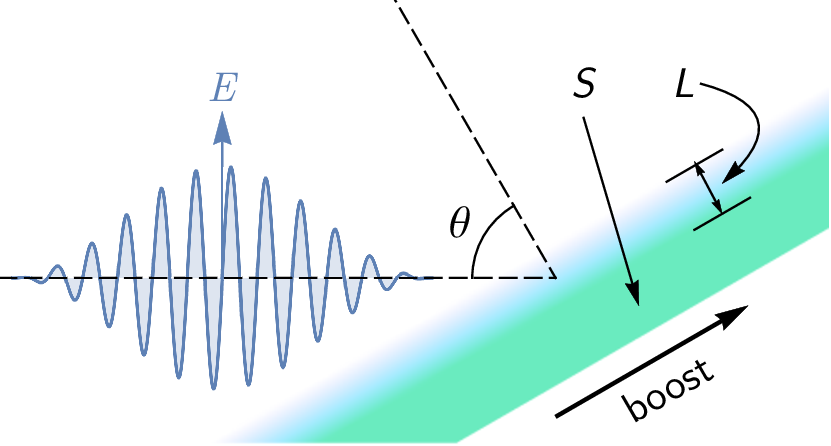}
	\caption{
		Interaction geometry: a $p$-polarized laser pulse is
		incident onto a plasma slab with bulk electron density
		determined by the $S$ parameter and linear density ramp
		with scale-length $L$.}
	\label{fig:Diagram}
	\end{figure}

The geometry under consideration is shown in \cref{fig:Diagram}.
In all the interaction scenarios we consider, the laser is $p$-polarized.
It is possible to simplify the problem to one with only a single spatial dimension
by boosting by $c\sin\theta$ in the direction parallel to the plasma
surface~\cite{Bourdier}. Neglecting the transverse intensity variation
of a focussed laser pulse, in this frame the laser may be treated as a
plane wave with electric-field amplitude $a_0 m c \omega \cos\theta/e$
and angular frequency $\omega \cos\theta$ which is normally incident on
a plasma with bulk density $n_e/\cos\theta$ and streaming velocity
$c \sin \theta$ perpendicular to the target normal.

The figures of merit we use to characterize XUV generation are: the effective
normalized amplitude $\aeff$, defined in \cref{eq:EffectiveA}; and the
conversion efficiency $\xce$, the fraction of the incident laser energy
re-emitted to light in the XUV frequency range (i.e. between wavelengths of
$10$ and $100\,\nm$) across the entire pulse train.
To assess the viability of using the RES regime as a
source of intense XUV radiation, we consider here plasmas with electron density
between 0.25 and 20 times the relativistic critical density and laser pulses
with intensity and duration that can achieved in currently-available laser
systems.
	
\section{XUV generation mechanism}
\label{sec:RES}

The key property of the laser-plasma interaction at near-critical
densities is that the electric-field amplitude of the re-emitted light can be
much larger than that of the incident light. One may see how
advantageous this is for XUV generation in the following way:
to keep the total energy of the pulse the same, an increase in
the electric-field amplitude by a factor $f = E_r/E_0$ must be accompanied
by a reduction in the duration by a factor $f^2$. The emission
of one such short pulse per optical cycle indicates that laser energy
is efficiently converted to high harmonics. Furthermore, the
fact that $a_0 \propto E/\omega$ raises the possibility of
generating relativistically intense XUV light, as the increase
in $\omega$ due to high harmonic generation may be partially
compensated by the electric-field amplitude increase that occurs at near-critical densities.

	\begin{figure}
	\centering
	\subfloat[]{\label{fig:ri-a}\includegraphics[width=\linewidth]{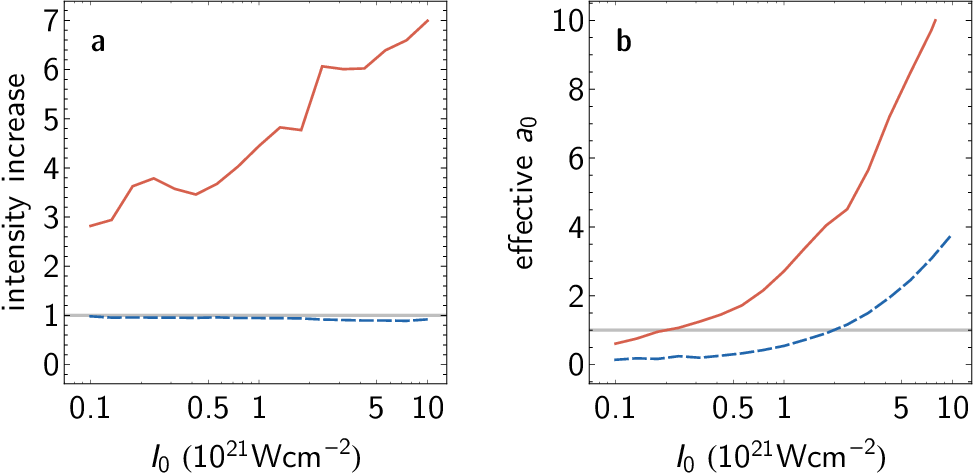}}
	\subfloat[]{\label{fig:ri-b}}
	\caption[Relativistic intensity]
			{a) The intensity increase and b) the effective $a_0$ of the
			re-emitted pulse when plasma with $S = 1$ (red) and $S = 20$
			(blue, dashed) is illuminated by a laser pulse with
			peak intensity $I_0$, duration $5\,\fs$ and wavelength
			$1\,\micron$ at $60^\circ$ to the target normal.
			}
	\label{fig:RelativisticIntensity}
	\end{figure}

Results of 1D PIC simulations using the spectral code \textsc{elmis}~\cite{Elmis}
that confirm this are shown in \cref{fig:RelativisticIntensity}.
For this set of simulations we consider laser pulses with lab-frame electric field
$(a_0 m c \omega/e) \sin\phi \cos^2(\phi/8)$ (for phase $|\phi| < 4\pi$)
and wavelength $1\,\micron$ incident at $\theta = 60^\circ$ onto
preplasma-free targets. The duration of such a pulse is $5\,\fs$
[full width at half maximum (FWHM) of the intensity profile].
The boosted frame coordinate system is defined such that
the laser phase $\phi = \omega\cos\theta (t' - z'/c)$ and that its electric
field vector lies along $x'$, corresponding to $p$-polarization.
The plasma has initial density $n_{e,0} = S a_0 \ncr / \cos\theta$,
following \cref{eq:S}, and streams along $x'$ with velocity
$v_{x',0}/c = -{\sin\theta}$. It has temperature 100~eV and
is represented by 200 macroelectrons and ions (C$^{6+}$, i.e. $Z/A = 0.5$) per
cell of size $\Delta z' = \lambda/(1800\cos\theta)$.

Let us compare the near and super-critical cases $S = 1$ and $S = 20$.
\Cref{fig:ri-a} shows that the peak intensity of the re-emitted pulse
is more than twice that of the incident light for all intensities in
the range $10^{20}$ to $10^{22}\,\Wcm^{-2}$ if the plasma density
satisfies $S = 1$. If instead $S = 20$, the peak intensity is always
bounded by that of the incident pulse. In both cases the re-emitted
radiation is broadband, though we will show later that the conversion
efficiency to XUV (harmonics from 10\textsuperscript{th} to
100\textsuperscript{th} order) is larger for lower $S$. First we
discuss how the intensity increase leads to enhanced $a_0$ for $S = 1$.

A electromagnetic wave is relativistically intense if it can
accelerate electrons from rest
to relativistic velocity within a single cycle. Given
the electric field of the re-emitted pulse $E_r$ as a
function of phase $\phi$, the transverse four-velocity of
an electron is given by
	\begin{equation}
	u_\perp = -{\int\! \frac{e E_r(\phi) \lambda}{2 \pi m c^2} \,\rmd\phi}.
	\end{equation}
We define the effective $a_0$ as
	\begin{equation}
	\aeff \equiv \max u_\perp.
	\label{eq:EffectiveA}
	\end{equation}
Here $\lambda$, the wavelength of the incident light,
appears for convenience because it defines the interval between
emission of individual attosecond pulses. Neither $u_\perp$ nor $\aeff$
depend directly on $\lambda$, because the phase $\phi$ contains a factor of $1/\lambda$.
The re-emitted pulse is relativistically intense in
the XUV range if $\aeff > 1$ when it is calculated
using an $E_r(\phi)$ that has been filtered to exclude
wavelengths outside the range $10\,\nm < \lambda < 100\,\nm$.
(For a monochromatic plane wave with electric-field amplitude
$a_0 m c \omega / e$, \cref{eq:EffectiveA} reduces to $a_0$ as
expected.)

We show the effective $a_0$ of the re-emitted XUV radiation in
\cref{fig:ri-b} as function of incident intensity. It is larger for
$S = 1$ than for $S = 20$ across the whole range of intensities.
For $S = 1$ the relativistic regime is reached for $I_0 > 2\times10^{20}\,\Wcm^{-2}$,
an order of magnitude less than that required if $S = 20$.
At $10^{22}\,\Wcm^{-2}$, the current intensity frontier~\cite{Hercules},
$\aeff$ reaches 10, raising the prospect of scaling relativistic
laser-plasma phenomena from optical to XUV wavelengths~\cite{Wettervik}.

	\begin{figure}
	\centering
	\subfloat[]{\label{fig:dyn-a}\includegraphics[width=\linewidth]{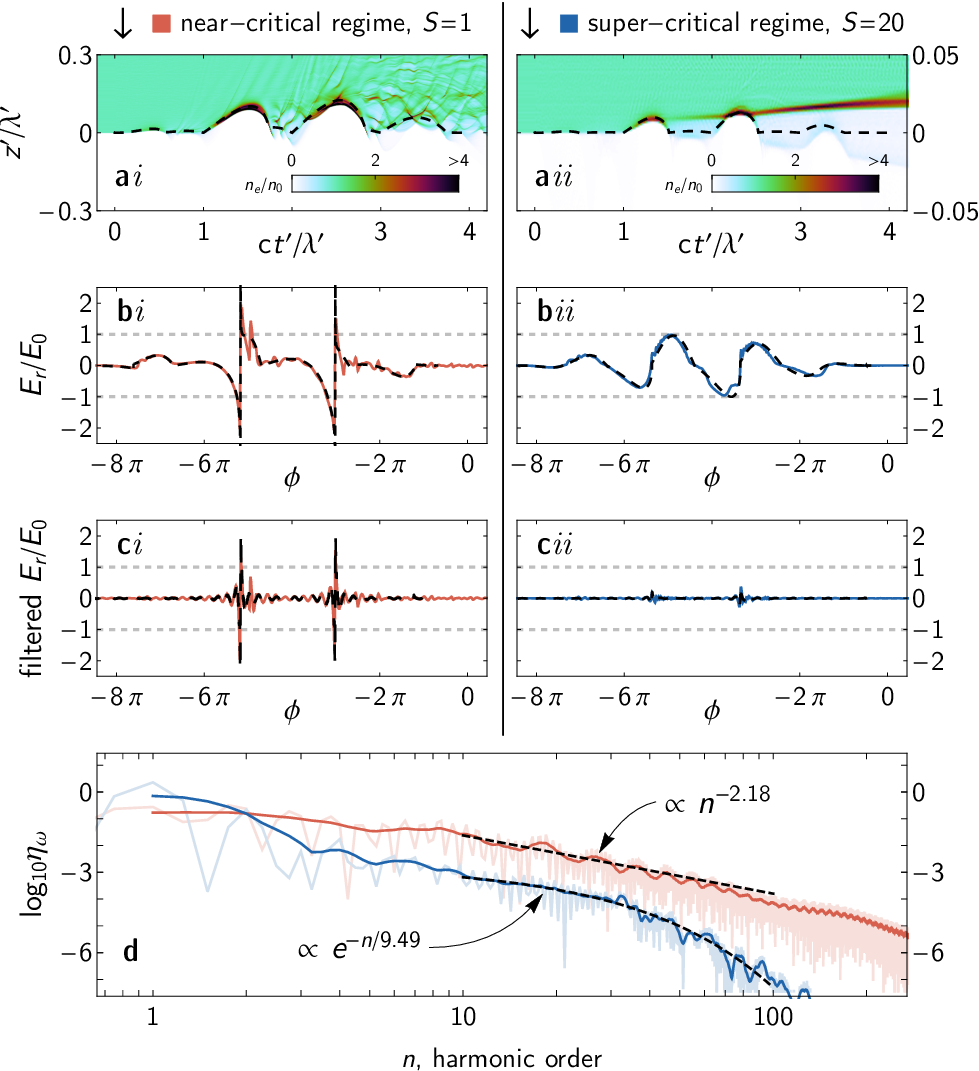}}
	\subfloat[]{\label{fig:dyn-b}}
	\subfloat[]{\label{fig:dyn-c}}
	\subfloat[]{\label{fig:dyn-d}}
	\caption[Plasma dynamics]
			{
			Comparison of XUV generation in laser illumination of
			near- and supercritical plasmas:
			a) the electron density in the boosted frame (colour scale)
			and the RES model prediction for the electron displacement
			(dashed);
			b) the lab-frame electric-field amplitude of the re-emitted light
			$E_r$ relative to the peak initial amplitude $E_0$ (colours)
			and RES model predictions of the same (black, dashed);
			c) as in b), but filtered to the XUV frequency range;
			and d) the spectral conversion efficiency from simulations:
			full (lighter colours), smoothed over
			the width of a single harmonic (darker colours), and
			analytic fits for the XUV frequency range (black, dashed).
			}
	\label{fig:NearSuper}
	\end{figure}

The origin of the intensity increase in the near-critical regime is
the increased amplitude of the plasma surface oscillation.
To demonstrate this, we compare the results of PIC simulations
of near-critical ($S = 1$) and supercritical ($S = 20$) plasmas
that are illuminated by a laser pulse with intensity
$I_0=10^{21}\,\Wcm^{-2}$ at an angle of incidence $\theta = 60^\circ$.
We also compare these findings to the theoretical predictions of
the RES model (for details and more benchmarking against simulations,
see \cite{Gonoskov2}).

\Cref{fig:dyn-a} compares the electron density in the boosted frame
as a function of axial displacement $z'$ and time $t'$, normalized
to the wavelength in the boosted frame $\lambda' = \lambda / \cos\theta$.
The oscillations arise because
the laser radiation pressure pushes electrons into the plasma, gathering
them into a thin sheet of high charge density at $z'_e$. The
uncompensated ion current in the region $0 < z' < z'_e$ sets up
electrostatic and magnetostatic fields, the former exerting
a force on the electron density spike that balances and then exceeds
the radiation pressure. The fact that this force is proportional
to the displacement $z'_e$, assuming the downstream region is entirely
cleared of electrons, is the origin of the name `relativistic electron
spring'~\cite{Gonoskov}.
We see that the RES equations predict both the scale and
period of the instantaneous displacement of the electrons very well.
At $S = 20$, a discrepancy arises towards larger $t'$ that is due to
ion motion; the breakup of the plasma surface leads to the last
half-wavelength of the pulse being reflected in a region of lower
density at $z'/\lambda' \simeq -0.01$.

During the phase of the motion when the electrons return towards
$z' = 0$, they acquire kinetic energy from the plasma fields, reaching
high $\gamma$. When their transverse velocity changes sign, at which
moment in the lab frame they propagate towards the observer along the specular
direction, a large-amplitude burst of radiation is emitted. This is
shown in \cref{fig:dyn-b}: for $S = 20$, the waveform acquires a
non-sinusoidal shape but the electric-field amplitude never exceeds the incident $E_0$,
whereas for $S = 1$ for the waveform is characterized by sharp transitions
between positive and negative field and amplitude increase
$\max(E_r/E_0) \simeq 2$.

This difference is clearly demonstrated when
the re-emitted light is filtered to the XUV frequency range (wavelengths
between 10 and $100\,$nm), as shown in \cref{fig:dyn-c}. We find that for
$S = 1$ the re-emitted attosecond pulses have electric-field amplitude twice
that of the incoming light, which corresponds to a peak intensity of
$4.0\times10^{21}\,\Wcm^{-2}$. For $S=20$, by contrast, the peak intensity
of the filtered pulses is $1.1\times10^{20}\,\Wcm^{-2}$.
We would expect $\aeff$ to be larger in the former case by a factor of
${\sim}6$ (the ratio of the pulse electric-field amplitudes); as \cref{fig:ri-b} shows
$\aeff$ to be $5$ times larger for $S = 1$, this is reasonably accurate.

As the RES equations predict the temporal evolution of the current
carried by the electron layer, we can use them to extract the
waveform of the re-emitted radiation; the theoretical and simulation
results are in good agreement for the total (\cref{fig:dyn-b}) and
spectrally filtered (\cref{fig:dyn-c}) electric-field amplitudes.
The overestimation of the peak value of $E_r$ for $S = 1$ is a
consequence of a singularity in the RES equations that occurs at
the instant the transverse velocity changes sign~\cite{Gonoskov2}.
This does not occur in the PIC simulations because the electron layer
is guaranteed to move with speed less than $c$.

The fact that the amplitude of the electric field, when filtered to
XUV frequencies, is larger for $S = 1$ than in $S = 20$ means that in
the former, there must be more energy carried in that frequency range
and therefore higher conversion efficiency. We can confirm this by
comparing the spectral conversion efficiencies (energy carried by harmonic
$n$ per unit energy of the incident pulse) we obtain from the simulations.
Note that the analytical fits we provide to the intensity spectrum
in the XUV frequency range deviate from the predictions of the standard ROM and
CSE models ($n^{-8/3}$ and $n^{-4/3}$ respectively); however,
power-law decay exponents from $-7/3$ to $-5/3$ have been reported,
as has their sensitivity to the choice of frequency range~\cite{Boyd}.
As the main purpose of our analysis is to find when the re-emitted XUV pulses can be
relativistically intense, the most important quantity is the
magnitude of the intensity spectrum; \cref{fig:dyn-d} shows that
the spectral conversion efficiency is at least an order of magnitude
greater for $S = 1$ than for $S = 20$ across the range $10 < n < 100$.
Specifically, we find that the total XUV conversion efficiency $\xce$,
obtained by integrating over this range, is 19.2\% for $S = 1$ and
0.67\% for $S = 20$.
The explains why the effective $a_0$ shown in \cref{fig:ri-b} exceeds
unity at a lower laser intensity if the plasma density is near-critical.

\section{Preplasma}
\label{sec:Preplasma}

Thus far we have considered interactions with plasma targets where the
bulk electron density is near-critical. This is experimentally changing,
but possible with the use of cryogenic, aerogel or porous foam targets~\cite{Fedeli}.
An alternative approach would be to exploit the finite contrast of a
high-intensity laser pulse, because prepulse causes heating and expansion
of the target and therefore the laser-plasma interaction takes place within
a density ramp with scale length $L$.
In this section we show that $L > 0$ can enhance XUV generation for
the same reason explored in \cref{sec:RES}, by reducing the apparent
density of the plasma.

To do so, we introduce an \emph{effective} $S$ parameter that
characterizes the laser-plasma interaction when $L$ is non-zero.
We take the preplasma to be a linear density
ramp with scale length $L$, i.e. $n_{e,0} = (S a_0 \ncr/\cos\theta) (1 + z'/L)$ for $-L < z' < 0$.
The laser pulse will penetrate the density ramp up to the point
$z' = z'_\text{max}$ where the electrostatic force of the charge
separation field is balanced against the magnetic forces arising
from the laser itself and the uncompensated ion current.
Assuming that all electrons in the region $z' < z'_\text{max}$
are swept forward, we find that
	\begin{equation}
	\frac{z'_\text{max}}{L} =
		-1 + \sqrt{\frac{\lambda (1 + \sin\theta)}{\pi L S}}
	\label{eq:PenetrationDepth}
	\end{equation}
where $S$ is calculated from the bulk plasma density and $\lambda$ is the
laser wavelength in the lab frame.

The value of $S$ that appears in \cref{eq:PenetrationDepth} is
calculated using the \emph{peak} normalized amplitude of the laser
pulse $a_0$.
The instantaneous penetration depth may be estimated by replacing
$S$ with its instantaneous value $S/\cos^2[\phi/(2N)]$,
where $N$ is the (arbitrary) number of cycles corresponding to the pulse duration.
Averaging the penetration depth over the interval $-N\pi < \phi < N\pi$
gives us an
average density which may be used to define an effective $S$
parameter for the interaction.
Assuming that the maximum penetration depth is smaller than the
preplasma scale-length $L$, we find
	\begin{align}
	\Seff
		&= \sqrt{\frac{4 S \lambda (1 + \sin\theta)}{\pi^3 L}},
	&
	L &> \frac{(1+\sin\theta) \lambda}{\pi S}.
	\label{eq:EffectiveS}
	\end{align}
For smaller $L$, the laser pulse breaks through the density ramp
and the bulk plasma is the source of the re-emitted radiation;
in this case we would have $\Seff = S$.

	\begin{figure}
	\centering
	\subfloat[]{\label{fig:ppd-a}\includegraphics[width=\linewidth]{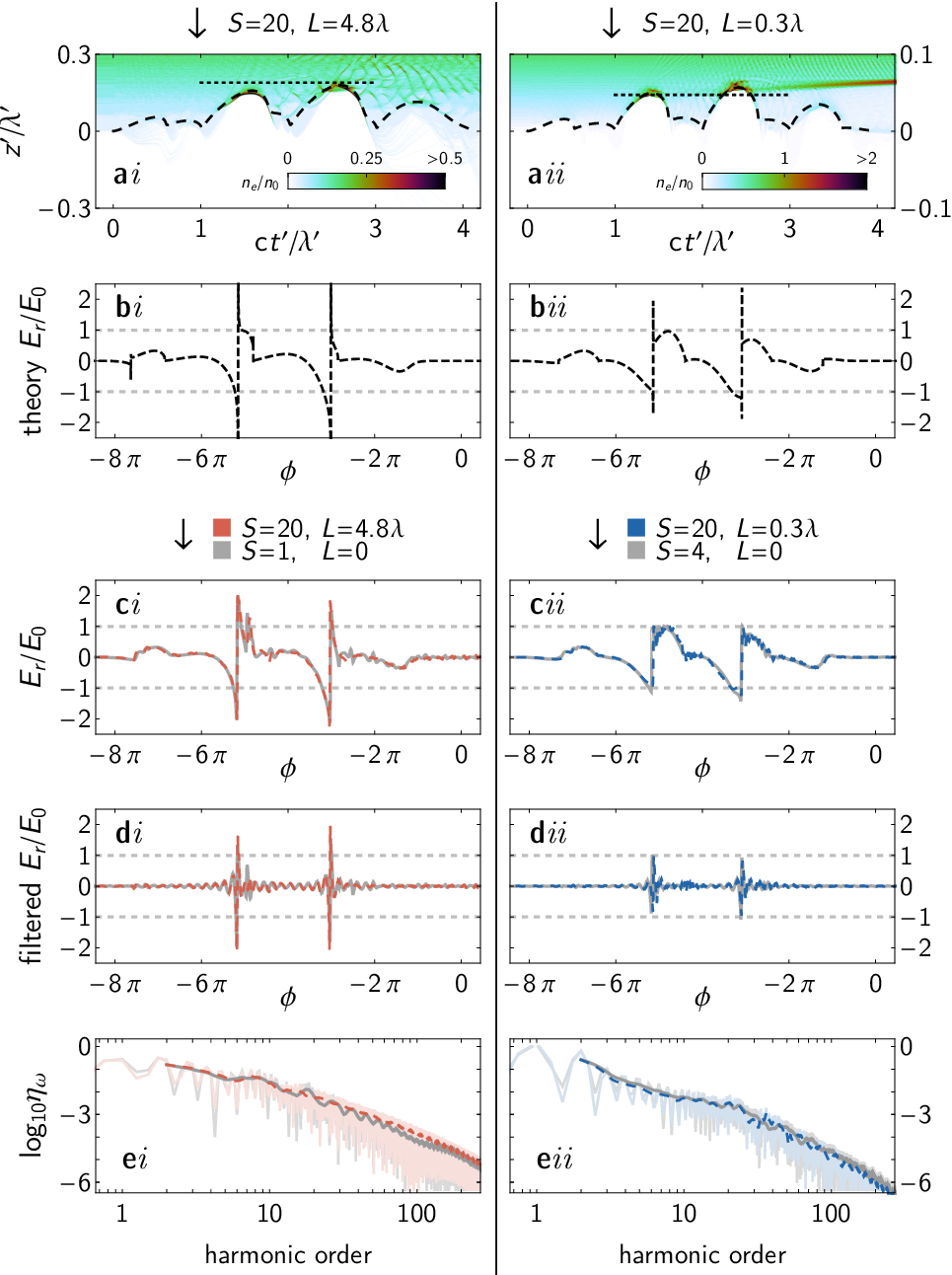}}
	\subfloat[]{\label{fig:ppd-b}}
	\subfloat[]{\label{fig:ppd-c}}
	\subfloat[]{\label{fig:ppd-d}}
	\subfloat[]{\label{fig:ppd-e}}
	\caption[Preplasma dynamics]
			{
			XUV generation in laser illumination of plasma with a linear density ramp:
			a) the electron density in the boosted frame (colour scale),
			the maximum displacment predicted by \cref{eq:PenetrationDepth} (dotted),
			and the RES model prediction of the electron displacement (dashed);
			b) the RES model prediction for the lab-frame electric-field amplitude
			$E_r$ (relative to the peak initial amplitude $E_0$);
			c) $E_r/E_0$ from simulation of a density ramp (colours), compared to the
			equivalent preplasma-free scenario (grey);
			d) as in b), but filtered to the XUV frequency range;
			and e) the spectral conversion efficiency: full (lighter colours)
			and smoothed over the width of a single harmonic (darker colours).
			}
	\label{fig:Preplasma}
	\end{figure}
	
This result leads us to expect the properties of
radiation emitted from a preplasma with given $\Seff$
to match those of radiation emitted from plasma
with a sharp density profile, if the density in the latter
case satisfies $S = \Seff$.
We demonstrate this for two exemplary cases, shown in
\cref{fig:Preplasma}. We consider plasma with bulk $S = 20$
and preplasma scale-lengths $L = 4.8$ and $0.3\,\micron$,
corresponding to $\Seff = 1$ and $4$ respectively.
The laser and other simulation parameters
are identical to those given in \cref{sec:RES}.

First we verify that, in both cases, the maximum displacement
of the electrons $z'_\text{max}$ is smaller than the preplasma
scale length. This can be seen in \cref{fig:ppd-a}, which shows
the electron density in the boosted frame as a function of time $t'$
and longitudinal displacement $z'$. We also find that our
simple prediction for $z'_\text{max}$ [\cref{eq:PenetrationDepth}] agrees
reasonably well with the simulation results;
furthermore, the instantaneous displacement of the plasma-vacuum
interface is modelled well by numerical solution of the RES
equations (as given by eq. 7 in \cite{Gonoskov2}).
The fact that the laser pulse is reflected within the density ramp
means that, as far as XUV generation is concerned, the interaction
occurs in a plasma of much lower density than the bulk value of
$S$ would imply.

The equivalence between the preplasma and preplasma-free cases
suggested by \cref{eq:EffectiveS} is evident when we compare
the waveforms of the emitted radiation.
\Cref{fig:ppd-c} shows excellent agreement between the electric-field
amplitudes obtained from simulation of these two scenarios,
provided that the density in the
preplasma-free case is chosen according to \cref{eq:EffectiveS}.
This is true even when the fields are filtered to the XUV
frequency range, as shown in \cref{fig:ppd-d}. (The results from the
preplasma-free case have been translated in time to aid the eye.)
The total $E_r$ predicted by solution of the RES equations for plasma
with a linear density ramp is shown in \cref{fig:ppd-b}. Aside
from the singularities that also affected the comparison in
\cref{fig:dyn-b}, the simulation results are reproduced well.

\Cref{fig:ppd-e} shows that
the spectral conversion efficiencies from simulation of preplasma
and equivalent preplasma-free targets agree well
across the XUV frequency range (harmonic orders between 10 and 100).
As a result, the total XUV conversion efficiencies $\xce$
are consistent between the preplasma and preplasma-free scenarios:
for a plasma slab with bulk density $S = 1$ (4), $\xce = 19.2\%$ (6.73\%),
whereas for a linear density ramp with $\Seff = 1$ (4), $\xce = 19.9\%$ (6.27\%).
Note that XUV generation is more efficient for lower $S$, as
we discussed in \cref{sec:RES}, and for lower $\Seff$.
Similarly, the peak electric-field amplitude of the re-emitted
radiation is larger for lower $S$ and for lower $\Seff$, leading
us to expect an increased effective $a_0$. We will discuss
the dependence of $\xce$ and $\aeff$ on the interaction parameters
in the next section.

\section{Optimal density, angle of incidence and preplasma scale-length}
\label{sec:1d}

We now discuss how laser-plasma interaction parameters may be chosen
to maximize the increase in reflected electric-field amplitude and the yield of high
harmonics. Previous work indicates that for single-wavelength light and
a perfectly sharp plasma-vacuum boundary, the optimal conditions are
$S \simeq 1$ and $\theta \simeq 60^\circ$~\cite{Gonoskov}. Here we extend
that analysis to the experimentally relevant scenario where there is
a preplasma in front of the bulk, with
parameter scans in laser intensity $I_0$,
plasma density $n_e$, angle of incidence $\theta$ and preplasma
scale-length $L$ using 1D boosted-frame PIC simulations using the
spectral code \textsc{elmis}~\cite{Elmis}.
The covered range is $10^{20}\,\Wcm^{-2} \leq I_0 \leq
10^{22}\,\Wcm^{-2}$, $0.25 \leq S \leq 16$, and $0^\circ < \theta < 80^\circ$.
The laser duration is increased to $15\,\fs$, i.e. its electric field
as a function of phase is $(a_0 m c \omega/e) \sin\phi \cos^2(\phi/24)$
for $|\phi| < 12\pi$.
Otherwise, the simulation parameters are as described in \cref{sec:RES}.

	\begin{figure}
	\centering
	\subfloat[]{\label{fig:1d-a}\includegraphics[width=\linewidth]{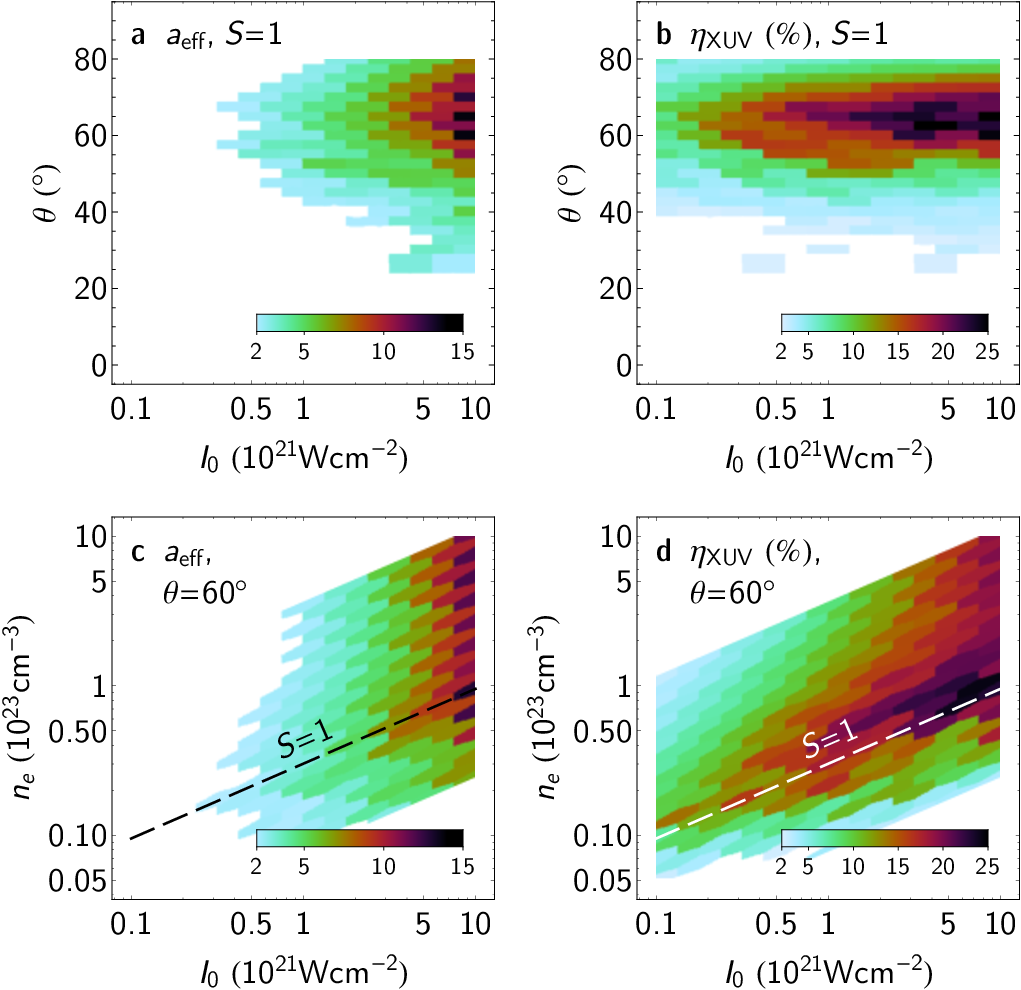}}
	\subfloat[]{\label{fig:1d-b}}
	\subfloat[]{\label{fig:1d-c}}
	\subfloat[]{\label{fig:1d-d}}
	\caption[Optimal target parameters]
			{
			(a,c) The effective $a_0$ of, and (b,d) the conversion efficiency
			to, XUV light
			when plasma with electron density $n_e = S a_0 \ncr$ is
			illuminated a laser pulse with peak intensity $I_0$,
			duration $15\,\fs$ and wavelength $1\,\micron$
			at an angle $\theta$ to the target normal.
			For (a,b) we fix $S=1$ and for (c,d) $\theta = 60^\circ$.
			$L = 0$ for all.
			}
	\label{fig:1d}
	\end{figure}

Let us first consider the situation where $S$ is fixed to be $1$, the
target is preplasma free, and the intensity and angle of incidence are
varied.
Here the electron density $n_e$ is implicitly increased with intensity
as required by \cref{eq:S}.
Results are given in \cref{fig:1d-a,fig:1d-b}.
We see that $\aeff > 2$ for intensities $\gtrsim 5\times10^{20}\,\Wcm^{-2}$
provided that the angle of incidence $\theta \simeq 60^\circ$. For angles
away from this optimum the electric-field amplitude increase associated with the
RES mechanism is less efficient and $\aeff$ falls. This can be partially
compensated by increasing the incident intensity; thus the acceptable
range of incidence angles broadens with increasing intensity.

We may strip out the dependence on incident intensity by considering the
conversion efficiency $\xce$, as this is normalized to the energy of the
incident laser pulse. \Cref{fig:1d-b} shows that $\xce$ does not demonstrate
a strong dependence on intensity for $I_0 > 5\times10^{20}\,\Wcm^{-2}$,
which is consistent with the $S$-scaling~\cite{GordienkoS}. There is little XUV emission
for $\theta \lesssim 20^\circ$ but more than 15\% of the pulse is converted
for $55^\circ \gtrsim \theta \lesssim 75^\circ$. This maximum
is consistent with that expected theoretically~\cite{Gonoskov}.

\begin{figure}
	\centering
	\subfloat[]{\label{fig:1dex-a}\includegraphics[width=\linewidth]{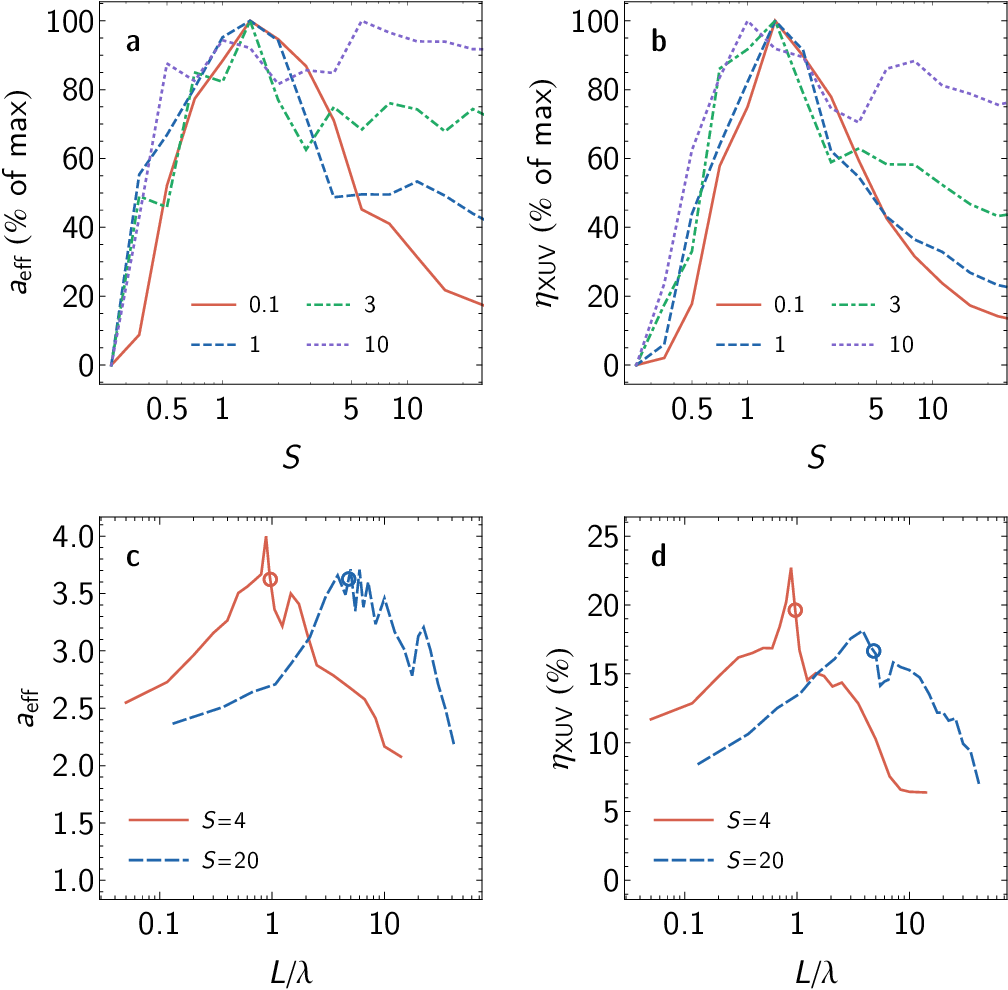}}
	\subfloat[]{\label{fig:1dex-b}}
	\subfloat[]{\label{fig:1dex-c}}
	\subfloat[]{\label{fig:1dex-d}}
	\caption[Optimal target parameters]
			{The a) effective $a_0$ and b) conversion efficiency to
			XUV when plasma is illuminated at $60^\circ$
			by a laser pulse with peak intensity $(0.1,1,3,10)\times10^{21}~\Wcm^{-2}$.
			Both $\aeff$ and $\xce$ have been scaled
			by the maximum value obtained at that intensity.
			The c) effective $a_0$ and d) conversion efficiency to
			XUV when plasma with preplasma scale length $L$ and
			bulk $S$ parameter is illuminated at $60^\circ$ by
			a laser pulse with peak intensity $10^{21}~\Wcm^{-2}$.
			Circles indicate the $L$ for
			which $\Seff = 1$ according to \cref{eq:EffectiveS}.}
	\label{fig:1dextra}
	\end{figure}

Fixing the angle of incidence at $60^\circ$, we explore intensity-density
space in \cref{fig:1d-c,fig:1d-d}. Increasing the plasma density at
constant intensity, i.e. increasing $S$, decreases $\aeff$ as we expect
given the results of \cref{sec:RES}. As the incident intensity increases,
so does $\aeff$ and the range of $S$ that ensures $\aeff > 2$.
The XUV conversion efficiency is largest for $S \gtrsim 0.5$ and decreases
as $S$ increases.
To clarify this point, we show the dependence of $\aeff$ and $\xce$
on increasing $S$ for fixed laser intensities $(0.1,1,3,10)\times 10^{21}~\Wcm^{-2}$
in \cref{fig:1dex-a,fig:1dex-b}, i.e. lineouts of \cref{fig:1d-c,fig:1d-d}
in the vertical direction. Both $\aeff$ and $\xce$ have been scaled
by the maximum value that is achieved at that particular intensity,
namely $\max(\aeff) = (0.77,3.5,6.8,13)$ and $\max(\xce) = (8.8,18,21,22)\%$.
The curves are qualitatively similar in that $\aeff$ and $\xce$ increase
as $S$ is increased up to $S \simeq 1$. Thereafter they begin to decrease,
with a gradient that becomes shallower at higher intensities. The range
of optimal $S$ broadens to the extent that, at $10^{22}~\Wcm^{-2}$, $\aeff$
is essentially constant for $S > 2$.
This violation of the $S$ similarity scaling is caused by ion motion,
which has been studied in the context of
increased broadening of high harmonics at normal incidence~\cite{Tang}.
While such effects will be present here, as the ion longitudinal
velocity and displacement are reduced at oblique incidence,
the principal effect is that breakup of the plasma surface is
enhanced, reducing the $S$ parameter accordingly.

In \cref{sec:RES} we saw that the presence of a density ramp in
front of the target lowers the effective $S$ encountered by the laser pulse,
permitting efficient XUV emission via the RES mechanism even for
plasma with bulk $S \gg 1$.
\Cref{fig:1dex-c,fig:1dex-d} shows that $\aeff$ and $\xce$
increase with increasing scale-length $L$ up to a certain point
and decrease thereafter. The open circles show that the locations
of these maxima are consistent with the $L$ for which $\Seff = 1$.
Furthermore the values of $\xce$ for these $L$, approximately 17\% for
both $S=4$ and $20$, are consistent with that which would be
obtained from an `ideal' target that has $S=1$ and is free of preplasma.
Over the range of $L$ shown, the normalized amplitude and
conversion efficiency are at least half this value. It is
evident that the largest allowable $L$ grows with increasing $S$,
as a longer density ramp is needed to reduce the effective $S$
parameter to the point that XUV generation is degraded. Taking
the lower bound for $\Seff$ to be ${\sim}0.25$, following
the results shown in \cref{fig:1d}, we can use
\cref{eq:EffectiveS} to estimate the largest allowable preplasma
scale length $L_\text{max} \simeq 3.8 S \lambda$ at $60^\circ$.
This is consistent with the ranges shown in \cref{fig:1dex-c,fig:1dex-d}.

This result demonstrates the robustness
of the RES mechanism and validates the use of an effective $S$
parameter to characterize the laser-plasma interaction. Expansion
of the target due to prepulse heating, if well-characterized and
controlled, can be an aid rather than an obstacle to XUV generation
by lowering the effective $S$ of the interaction.

\section{Divergence, brilliance and intensity}
\label{sec:2d}

To determine the angular divergence and brilliance of the XUV pulse so generated,
we have carried out a parameter scan over $0.5 \leq S < 15$ and $40^\circ \leq \theta \leq 75^\circ$
at fixed intensity $I_0 = 5\times10^{21}\,\Wcm^{-2}$ using 
the 2D3V PIC code \textsc{epoch}~\cite{Epoch}.
For all simulations described in this section, the laser is plane-polarized and
focussed to a spot with waist $w_0 = 4\,\micron$ at the plasma surface.
Its temporal profile and wavelength are the same as in the 1D simulations:
$\cos^2$ with FWHM duration $15\,\fs$ and $\lambda = 1\,\micron$ respectively.
The simulation domain, $[0,16]\times[-3,24]\lambda$ in the $x$ and $z$-directions,
contains electron-ion plasma ($Z/A = 0.5$) in the region $z \leq 0$ with
initial temperature $100\,\eV$.
The resolution varies between 200 and 500 cells per $\lambda$, and the
number of particles per cell between 24 and 48 for each species.
The laser is injected from the left boundary at an angle $\theta$ to the target normal;
to reduce the length of plasma, and so the number of computational particles,
required to model the interaction, the simulation domain moves along $x$
with velocity $c\sin\theta$.
The electric-field amplitude and harmonic content of the reflected light is analyzed once
the pulse has moved a perpendicular distance of $12\,\micron$ from the plasma
surface. Preplasma is added in the same way as for the 1D simulations, by
including a linear density ramp of length $L$ in front of the bulk plasma.

In moving from one to two spatial dimensions we encounter two new physical
effects that alter the XUV generation process. The first is that the conservation
of transverse kinetic momentum, which applies exactly in 1D, is lifted, leading
to increased electron heating and reduction of the plasma reflectivity.
Secondly, the variation in laser intensity across the plasma surface (or, equivalently,
the variation in the effective $S$) leads to a spatially varying displacement
of the electron-ion boundary and to wavefront curvature of the reflected light.
Both of these effects are more pronounced for plasma of lower $S$.

	\begin{figure}
	\subfloat[]{\label{fig:snapshot-a}\includegraphics[width=\linewidth]{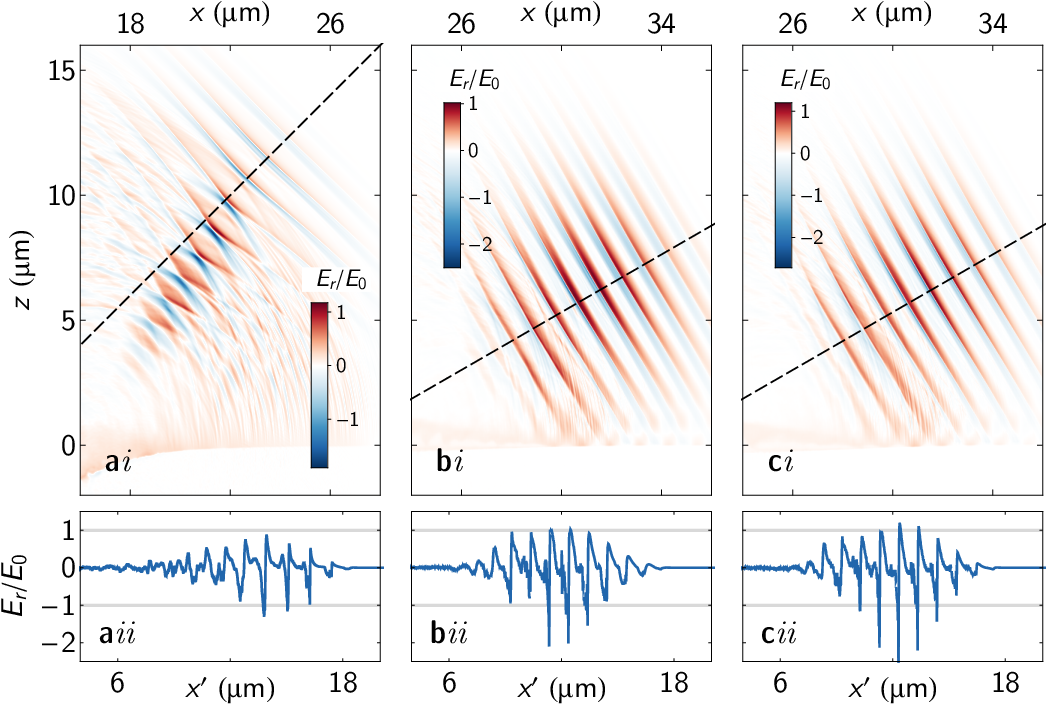}}
	\subfloat[]{\label{fig:snapshot-b}}
	\subfloat[]{\label{fig:snapshot-c}}
	\caption[Pulse snapshots]{
			(upper) Colour maps and (lower) lineouts along the specular
			direction of the electric-field amplitude of the re-emitted pulse when
			plasma with bulk $S$ and preplasma scale-length $L$ is
			illuminated at angle $\theta$ by a laser pulse with peak intensity
			$5\times10^{21}\,\Wcm^{-2}$,
			where a) $S = 0.5$, $L = 0$, $\theta = 45^\circ$;
			b) $S = 4$, $L = 0$, $\theta = 60^\circ$; and
			c) $S = 4$, $L = 0.3\,\micron$, $\theta = 60^\circ$.}
	\label{fig:snapshot}
	\end{figure}
	
In \cref{fig:snapshot} we show the re-emitted pulse electric-field amplitude from 2D simulations
for various $S$ and $\theta$. For the lowest density, \cref{fig:snapshot-a}, the
wavefronts of the leading edge are almost flat and they travel along the specular
direction. Behind this, however, the wavefronts become strongly distorted as the
pulse reflects from a curved plasma surface. The lineout along the
specular direction shows that the peak electric-field amplitude is approximately equal to that
of the incident pulse, so the amplitude increase we expect from the RES
mechanism is lost.

These effects are mitigated by moving to higher $S$. For $S = 4$,
\cref{fig:snapshot-b} shows that the wavefronts are flat across the entire
pulse and the peak electric-field amplitude is twice that of the incident pulse. If a
linear density ramp with scale-length $0.3\,\micron$ is introduced in front
of the bulk plasma, as shown in \cref{fig:snapshot-c}, we find that the peak
electric-field amplitude is even larger, at $2.7\times$ the initial value.
This corresponds achieving a peak intensity of
$3.6\times10^{22}\,\Wcm^{-2}$, above the current record~\cite{Hercules}.
It should be noted that this increase arises in absence of any focussing, as
the plasma surface has not been specially curved
nor is there appreciable ponderomotive denting.
It has been proposed that shaped plasma targets may be used to focus high harmonics
to extreme intensity, exploiting the reduction in the diffraction
limit~\cite{Naumova,GordienkoFocussing}; such effects would stand
in addition to the intensity increase arising from the RES mechanism that
we show here.

	\begin{figure}
	\centering
	\subfloat[]{\label{fig:2d-a}\includegraphics[width=\linewidth]{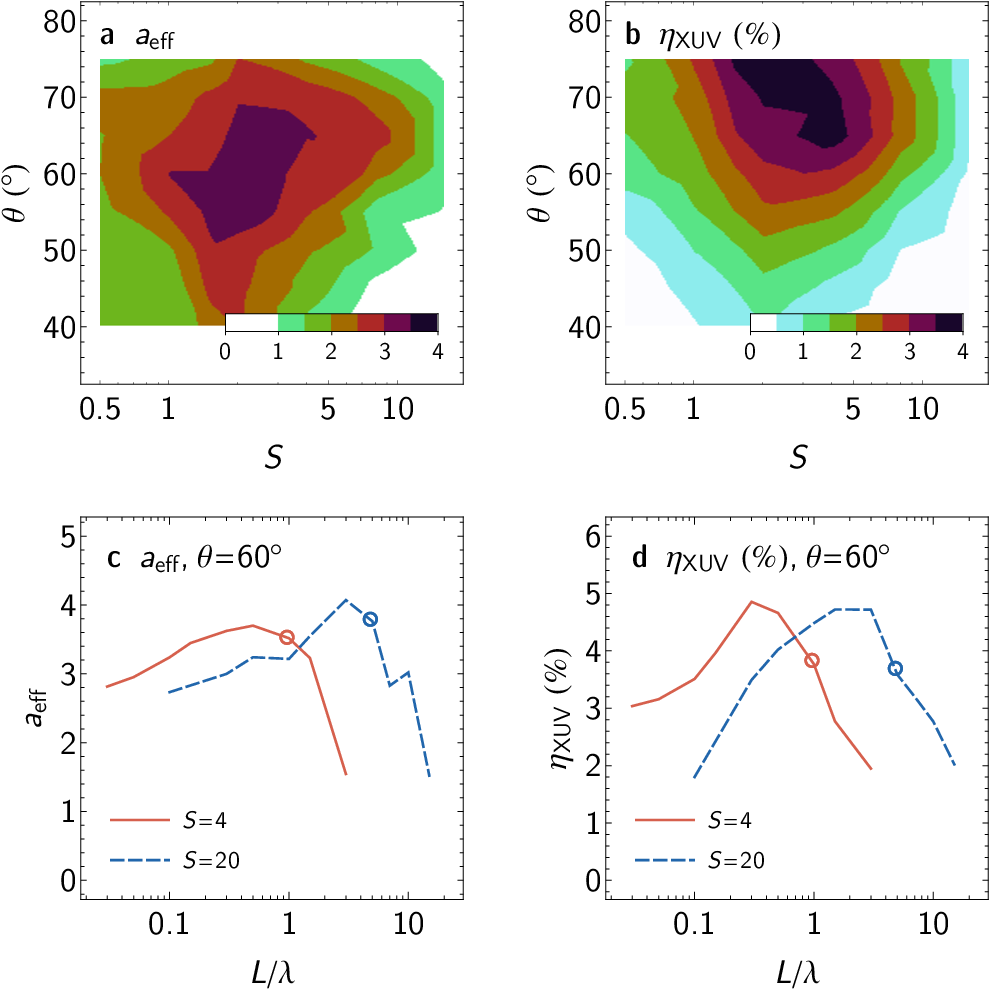}}
	\subfloat[]{\label{fig:2d-b}}
	\subfloat[]{\label{fig:2d-c}}
	\subfloat[]{\label{fig:2d-d}}
	\caption[2d]
			{a) The effective $a_0$ and b) the conversion efficiency
			in the XUV frequency range
			when plasma with electron density $n_e = S a_0 \ncr$ is
			illuminated by laser light with peak intensity
			$5\times10^{21}\,\Wcm^{-2}$ at angle $\theta$ to the
			target normal.
			c) The effective $a_0$ and d) XUV conversion efficiency
			when plasma with linear density ramp (scale-length $L$)
			and bulk $S = $~(blue, solid)~4 and (yellow, dashed)~20
			is illuminated at $60^\circ$ by a laser pulse with
			peak intensity $5\times10^{21}\,\Wcm^{-2}$.
			Circles indicate the $L$ for which $\Seff = 1$,
			as predicted by \cref{eq:EffectiveS}.
			}
	\label{fig:2d}
	\end{figure}

The consequence of increased electron heating and non-uniformity of the plasma
surface is to shift the optimum for XUV generation from $S = 1$ to $S \simeq 4$,
and reduce both $\aeff$ and the conversion efficiency $\xce$.
Nevertheless, the dependence of the former on $S$ and $\theta$ in the 2D case,
shown in \cref{fig:2d-a}, demonstrates qualitative agreement with that
found with 1D simulations. The reflected pulse is relativistically intense
in the XUV frequency range for a broad range of angles around $60^\circ$
and $S \simeq 3$.
Similarly, the XUV reflectivity shown in \cref{fig:2d-b} still has the broad peak
between $55^\circ$ and $75^\circ$ we expect based on 1D theory and simulation.
The best value of $4\%$ is approximately a factor of four smaller than that
obtained in 1D. We attribute this to increased electron heating caused by
wavelength-sized density perturbations, which are induced along the plasma
surface by the oblique incidence of the laser pulse.

The final result of \cref{sec:1d}
was that the effective $a_0$ and XUV conversion efficiency of plasmas with $S$ above the
optimum could be improved by introducing a preplasma of sufficient length that
the peak of the incoming pulse reflects off a surface with lower effective
density. \Cref{fig:2d-c,fig:2d-d} show that this principle still applies:
the effective $a_0$ may be increased to $\sim4$ and $\xce$ to 5\%
for both $S=4$ and $S=20$. The preplasma scale-length required is smaller than
that which would make $\Seff = 1$ according to \cref{eq:EffectiveS};
this is explained by the fact that in 2D, the optimal $S$ is larger and so
the optimal $L$ must be reduced. Nevertheless, the qualitative behaviour is
as we find in 1D. We find that the generation of relativistically intense XUV light with
good conversion efficiency is robust against the presence of preplasma over
a wide range of scale-lengths.

	\begin{figure}
	\centering
	\subfloat[]{\label{fig:br-a}\includegraphics[width=\linewidth]{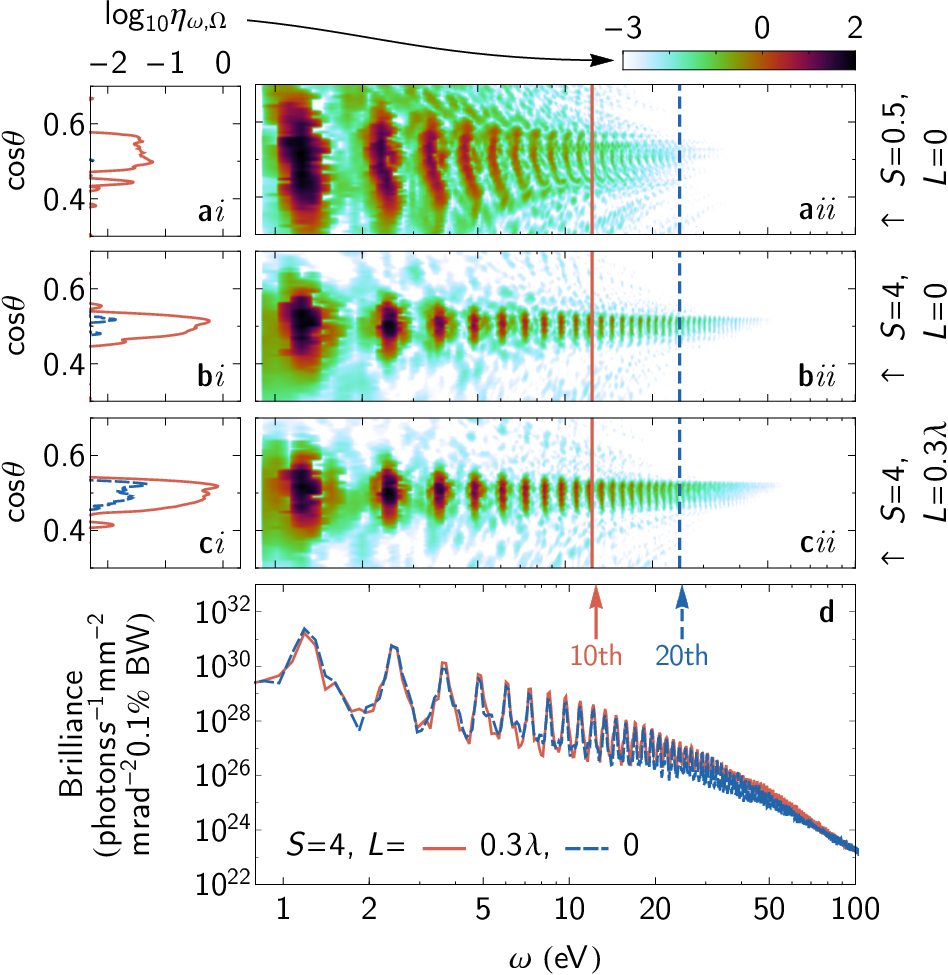}}
	\subfloat[]{\label{fig:br-b}}
	\subfloat[]{\label{fig:br-c}}
	\subfloat[]{\label{fig:br-d}}
	\caption[Spectral conversion efficiency and brilliance]
			{
			(a--c) $\spR$, the conversion efficiency per unit photon energy $\omega$
			per unit solid angle $\cos\theta$, for plasma with
			given $S$ and preplasma scale length $L$ irradiated
			by a laser pulse with peak intensity $5\times10^{21}\,\Wcm^{-2}$
			at incidence angle $60^\circ$,
			showing slices through
			colour maps \textit{ii}) at the
			(blue, solid) 10th and (yellow, dashed) 20th harmonics.
			(d) The brilliance for $S = 4$ and given $L$ at $60^\circ$.
			}
	\label{fig:brilliance}
	\end{figure}

The flat wavefronts shown in \cref{fig:snapshot-b,fig:snapshot-c} suggest
that the emission of high harmonics is closely collimated along the
specular direction, with angular divergence comparable to that of the incoming
laser pulse. We show in \cref{fig:br-a,fig:br-b,fig:br-c} colour maps of $\spR$, the
conversion efficiency per unit frequency $\omega$ and solid angle $\cos\theta$. The
harmonics become more distinct, with smaller divergence, as $S$ increases.
Slices at the 10\textsuperscript{th} and 20\textsuperscript{th} harmonics
show that for $S = 4$ the reflected energy is concentrated
within a window $\Delta(\cos\theta) \simeq 0.05$, corresponding to a half-angle
of $3^\circ$ at $60^\circ$. For comparison, the incident laser pulse with waist
$w_0 = 4\,\micron$ and wavelength $1\,\micron$ has half-angle divergence
$\theta_\text{laser} = \lambda/(\pi w_0) \simeq 5^\circ$. One might expect
that the wavelength dependence in this relation leads to decreasing
divergence with increasing harmonic order. However \cref{fig:brilliance} shows
that the divergence of the 20\textsuperscript{th} harmonic is near,
not half, that of the 10\textsuperscript{th}, and therefore the dominant
effect is the overall divergence of the laser pulse.
The distribution of energy in $\omega$ and $\cos\theta$ is broader and
noisier for $S = 0.5$ than it is for $S=4$. The broadening is caused
by ponderomotive denting of the plasma surface, which increases the curvature
of the reflected wavefronts, i.e. emission
of high-frequency radiation away from the specular axis. \Cref{fig:br-b}
and \cref{fig:br-c} differ only by the inclusion of preplasma, which
does not affect the angular properties of the radiation
aside from an overall increase in the total conversion efficiency.

Finally, we use the spectral conversion efficiency from simulation to estimate the
brilliance, which is a measure of photon phase-space density. Assuming a
fiducial distance in the $y$-direction (perpendicular to the simulation plane)
of $4\,\micron$ to determine the total energy of the pulse, a
spot size of $\pi w_0^2$, a duration equal to the laser pulse of 15~fs,
and a half-angle divergence of $3^\circ$, we show in \cref{fig:brilliance}
that the brilliance is of order $10^{23}$~photons/s/mm$^2$/mrad$^2$
(0.1\% bandwidth) at $\omega = 100\,\eV$, for the optimal parameters of
$S = 4$, $\theta = 60^\circ$ and a linear density ramp of preplasma
with length $L = 0.3\,\micron$.
While this is comparable to the brilliance
achieved in third-generation synchrotron light sources~\cite{ALS},
it is lower than that achieved by advanced gas harmonic
sources~\cite{Popmintchev}, and six to nine orders of magnitude
smaller than that reached by X-ray free electron lasers~\cite{Ackermann}.
The benefit of employing an ultraintense laser-plasma interaction
to generate XUV radiation is that the larger pulse energy and femtosecond
duration make it possible for the high harmonics to be
both bright and relativistically intense, a unique capability
of the near-critical interaction scenario presented here.

\section{Conclusions}
\label{sec:conc}

We have explored how the relativistic electron spring mechanism
leads to bright, intense bursts of XUV radiation when plasma with electron
density $n_e$ satisfying $1 < S = n_e/(a_0 \ncr) < 10$ is illuminated by intense
laser light ($a_0 \gg 1$). The physical origin of this enhanced emission is
the storage of energy in plasma electromagnetic fields when the electron-ion
boundary is displaced by the oscillating radiation pressure of the laser.
We have justified the theoretical prediction of the optimal parameters
$S = 1$, $\theta = 60^\circ$ with high-resolution 1D PIC simulation
and shown further that the presence of preplasma can be beneficial
by lowering the effective $S$ of a plasma with bulk density higher than
would be optimal.

Parametric analysis with 2D simulations indicates that electron
heating and ponderomotive denting of the plasma surface shift
the optimal $S$ upwards to $S \simeq 4$.
However, the consequent reduction in conversion efficiency
can be partially offset by preplasma, permitting conversion efficiencies
of ${\sim}5\%$ to XUV light collimated at the degree-level even
for $S = 20$.
The re-emitted pulse can reach a peak intensity five
times greater than the incident pulse, sufficient to be relativistically
intense when filtered to the XUV frequency range. This capability
provides opportunities for XUV-pump--XUV-probe experiments, exploiting
the ultrashort duration of the pulse and the comparatively compact
size of a high-intensity laser facility.

\begin{acknowledgements}
The authors acknowledge support from
the Knut and Alice Wallenberg Foundation (T.G.B., A.G., M.M.),
the Swedish Research Council (grants 2013-4248 and 2016-03329, M.M., and grant 2017-05148, A.G.),
the Russian Science Foundation (project no. 16-12-10486, A.G., RES calculations)
and the Russian Foundation for Basic Research (project no. 15-37-21015, A.G.).
Simulations were performed on resources provided by the Swedish National
Infrastructure for Computing (SNIC) at the High Performance Computing
Centre North (HPC2N).
\end{acknowledgements}

\bibliography{references}

\end{document}